\documentclass[conference]{IEEEtran} 
\usepackage[utf8]{inputenc}
\usepackage[pdftex]{graphicx}
\usepackage{todonotes}
\usepackage{balance}

\newif\ifauthor
\authortrue

\usepackage[english]{babel}

\usepackage[%
rm={oldstyle=false,proportional=true},%
sf={oldstyle=false,proportional=true},%
tt={oldstyle=false,proportional=true,variable=true},%
qt=false%
]{cfr-lm}

\usepackage{graphicx}

\usepackage{paralist}

\usepackage[T1]{fontenc}

\usepackage[math]{blindtext}

\usepackage{csquotes}

\usepackage{microtype}

\usepackage{url}
\makeatletter
\def\Url@twoslashes{\mathchar`\/\@ifnextchar/{\kern-.2em}{}}
\g@addto@macro\UrlSpecials{\do\/{\Url@twoslashes}}
\makeatother
\makeatletter
\g@addto@macro{\UrlBreaks}{\UrlOrds}
\makeatother

\usepackage{xcolor}

\usepackage{scrextend}
\usepackage{footnote}
\usepackage[bottom]{footmisc}

\usepackage[
bookmarks=false,
breaklinks=true,
colorlinks=true,
linkcolor=black,
citecolor=black,
urlcolor=black,
pdfpagelayout=SinglePage,
pdfstartview=Fit
]{hyperref}
\usepackage[all]{hypcap}

\usepackage[capitalise,nameinlink]{cleveref}
\crefname{section}{Sect.}{Sect.}
\Crefname{section}{Section}{Sections}

\usepackage{xspace}

\newcommand{\etal}{~et al.\ }

\usepackage{array}
\newcolumntype{L}[1]{>{\raggedright\let\newline\\\arraybackslash\hspace{0pt}}m{#1}}
\newcolumntype{C}[1]{>{\centering\let\newline\\\arraybackslash\hspace{0pt}}m{#1}}
\newcolumntype{R}[1]{>{\raggedleft\let\newline\\\arraybackslash\hspace{0pt}}m{#1}}
\newcommand{\cell}[2][c]{\begin{tabular}[#1]{@{}c@{}}#2\end{tabular}}
\DeclareFontFamily{U}{MnSymbolC}{}
\DeclareSymbolFont{MnSyC}{U}{MnSymbolC}{m}{n}
\DeclareFontShape{U}{MnSymbolC}{m}{n}{
    <-6>  MnSymbolC5
   <6-7>  MnSymbolC6
   <7-8>  MnSymbolC7
   <8-9>  MnSymbolC8
   <9-10> MnSymbolC9
  <10-12> MnSymbolC10
  <12->   MnSymbolC12%
}{}
\DeclareMathSymbol{\powerset}{\mathord}{MnSyC}{180}

\usepackage{listings}
\usepackage{multirow}
\usepackage{listings}
\usepackage{amsmath}
\usepackage{pgf}
\usepackage{tikz}
\usetikzlibrary{arrows,automata}
\usepackage[ruled]{algorithm2e}
\usepackage{cite}
\def\etal.{et~al.}

\def\baselinestretch{0.985}

\graphicspath{{.}}

\begin{document}

\input glyphtounicode.tex
\pdfgentounicode=1

\title{Securing Conditional Branches in the Presence of Fault Attacks}
\ifauthor
    \author{
        \IEEEauthorblockN{Robert Schilling\IEEEauthorrefmark{1}\IEEEauthorrefmark{2}, Mario Werner\IEEEauthorrefmark{1}, Stefan Mangard\IEEEauthorrefmark{1}}
        \IEEEauthorblockA{\IEEEauthorrefmark{1}Graz University of Technology
        \\firstname.lastname@iaik.tugraz.at}
        \IEEEauthorblockA{\IEEEauthorrefmark{2}Know-Center GmbH\vspace{-4ex}}
    }
    \date{September 2017}
\fi

\maketitle

\begin{abstract}
In typical software, many comparisons and subsequent branch operations are highly critical in terms of security. Examples include password checks, signature checks, secure boot, and user privilege checks. For embedded devices, these security-critical branches are a preferred target of fault attacks as a single bit flip or skipping a single instruction can lead to complete access to a system. In the past, numerous redundancy schemes have been proposed in order to provide control-flow-integrity (CFI) and to enable error detection on processed data. However, current countermeasures for general purpose software do not provide protection mechanisms for conditional branches. Hence, critical branches are in practice often simply duplicated.

We present a generic approach to protect conditional branches, which links an encoding-based comparison result with the redundancy of CFI protection mechanisms. The presented approach can be used for all types of data encodings and CFI mechanisms and maintains their error-detection capabilities throughout all steps of a conditional branch. We demonstrate our approach by realizing an encoded comparison based on AN-codes, which is a frequently used encoding scheme to detect errors on data during arithmetic operations. We extended the LLVM compiler so that standard code and conditional branches can be protected automatically and analyze its security. Our design shows that the overhead in terms of size and runtime is lower than state-of-the-art duplication schemes.
\end{abstract}


\section{Introduction}
\label{sec:Introduction}
A conditional branch determines the program flow of the executed software based on a flag or based on the comparison of two values. While the basic functionality of a conditional branch is quite simple, the correct execution is highly critical for the security of computer systems. In the end, it is a conditional branch that decides whether or not an entered password is considered correct by a system, a system update is performed, a signature check is considered successful, or a user is granted access to a privileged function. The security implications are huge if a critical program flow decision is not taken correctly and for example unauthenticated software is executed~\cite{riscure2017}.

Under normal conditions, conditional branches execute correctly, i.e.,~the branch is performed according to the comparison. However, there exist fault attacks, which allow an attacker to change the state of a system by using, i.e.,~a laser~\cite{DBLP:conf/fdtc/WoudenbergWM11} or by manipulating the clock signal or the supply voltage~\cite{DBLP:conf/fdtc/BarenghiBPP09}.
The effects of fault induction can include the skipping of instructions~\cite{DBLP:conf/fdtc/SchmidtH08}, the redirection of memory accesses, or flipping or forcing bits in memory or registers. Even bypassing secure boot mechanisms is possible~\cite{riscure2017}.

The research field of studying fault inductions on the security of cryptographic algorithms is very active~\cite{DBLP:conf/acisp/ChenY03,DBLP:conf/wistp/TunstallMA11}. Techniques exist to reveal the keys of different cryptographic functions for many different fault models~\cite{DBLP:journals/joc/BonehDL01,DBLP:conf/crypto/BihamS97}. There also exist several proposals for countermeasures~\cite{DBLP:conf/wistp/KimQ07, DBLP:conf/fdtc/MalkinSY06}.
When it comes to attacks and countermeasures for the general execution of software, there are significantly fewer publications. Countermeasures that have been proposed so far for securing software execution can be grouped into two classes.
First, there are countermeasures that aim for ensuring control-flow integrity (CFI) in the setting of fault attacks, like~\cite{DBLP:conf/cardis/WernerWM15}. Concerning conditional branches, these countermeasures only ensure that one of the two possible execution paths and no completely different path is taken after a conditional branch. However, these CFI countermeasures do not protect the decision which path is taken against fault attacks. The second class of countermeasures are redundancy mechanisms for data~\cite{hamming1950error, DBLP:conf/dsn/WangKK09}. For example,~\cite{DBLP:conf/safecomp/FetzerSS09} shows how to protect variables during arithmetic operations using AN-codes. However, as also pointed out in~\cite{DBLP:conf/hase/HoffmannUDSLS14}, such schemes only protect data values and their processing and no branching operations based on the data.

Today, there is a gap. There exist CFI and data protection schemes against fault attacks. However, there is no method of linking them efficiently such that the decision on which execution path to take is protected at the same level as the control-flow and the processing of the data. In practice, a method to avoid this gap is to do not only one conditional branch, but to check the condition for the branch again after the branch has been taken. This duplication approach increases security and can be scaled to an arbitrary order. However, this duplication approach leads to significant overheads on the one side, and it can be attacked by inducing multiple times the same fault. The options for creating diversity by using different branches to make attacks harder is limited. 
Typically, it is the same hardware multiplexer for all branches, which decides which address is loaded next and remains as a single point of failure. In this article, we close the existing gap by providing protection for conditional branches which is encoding-based, like the redundancy schemes for data and CFI. Our concrete contributions are as follows:
\begin{enumerate}
	\item We present a generic solution that closes the gap of unprotected conditional branches in the presence of a CFI protection scheme. Conditional branches are protected by linking a redundant comparison result with the redundancy of the CFI protection scheme.
	\item We show that we can use AN-codes to efficiently perform a redundant comparison of encoded values which preserves the redundancy.
	\item We present an LLVM compiler extension to automatically identify and protect conditional branches based on the concept of AN-codes. We provide experimental results showing that the overhead in terms of code size and runtime is lower than state-of-the-art duplication schemes. Furthermore, a bootloader application can efficiently be protected with 2.4\% code overhead and with less than 0.1\% runtime overhead.
\end{enumerate}

The structure of this paper is as follows. 
We define the threat model and discuss existing countermeasures against faults in Section~\ref{sec:background}. 
In Section~\ref{sec:Protected_Conditional_Branches}, we present how we protect conditional branches in this setting. 
We discuss a novel approach to compute a redundant comparison result used for protected conditional branches in Section~\ref{sec:AN-Code_Comparisons}.
In Section~\ref{sec:Implementation_Evaluation}, we present a compiler extension to protect conditional branches automatically and evaluate the overhead. Section~\ref{sec:analysis} analyzes the security of the countermeasure, and finally, in Section~\ref{sec:Conclusion} we conclude this paper.

\section{Fault Protected Software Execution}
\label{sec:background}

Throughout this paper, we consider an attacker with physical access to an embedded system. Hence, the attacker can tamper with the surroundings to induce faults, which can modify data, code, and also signals like the comparison result. Faults can occur once or multiple times with multiple bits modified. We assume the presence of an instruction-granular CFI protection scheme, protecting the execution of instructions and the selection of the operands.
Furthermore, we assume the data to be encoded redundantly.

\subsection{Code Protection with Control-Flow Integrity}
\label{sec:cfi}

CFI protection schemes~\cite{DBLP:conf/ccs/AbadiBEL05} typically protect the execution of code against software attacks. However, recent work extends these countermeasures in the context of fault attacks, where they try to protect the sequence of basic blocks or even the sequence of instructions~\cite{DBLP:conf/date/ClercqKC0MBPSV16, DBLP:conf/cardis/WernerWM15, DBLP:journals/corr/SullivanAGDSJ17}.\looseness=-1 

CFI protection schemes in the context of fault attacks rely on an internal state $S$, which is modified by each executed instruction. 
Independent of the concrete CFI protection scheme, control-flow transfers like conditional branches require special treatment. 
On control-flow transfers, the internal state $S$ diverges, because the instructions diverge. 
When the control-flow graph merges at a later point in the program, the CFI state $S$ also needs to merge. To support this, CFI protection schemes either use correction values or replace the state.

Although CFI protection mechanisms can deal with conditional branches, they can not protect them. Such a scheme only ensures that one of the correct successor'	s blocks is executed after the branch, but the correct selection is completely unprotected leaving a single point of failure.

\subsection{Data Protection with AN-Codes}
\label{sec:an_code_primer}

To counteract soft errors  in modern CMOS devices caused by radiation, encoding schemes like~\cite{hamming1950error, DBLP:conf/dsn/WangKK09} have been proposed. However, these mechanisms are developed for a protected storage in the memory not for protecting operations inside the CPU.
We focus on AN-codes~\cite{DBLP:journals/tc/Brown60}, which are well suited for fault protection~\cite{DBLP:conf/safecomp/FetzerSS09}, and natively supports different arithmetic operations.
AN-codes have the form $n_c = A\cdot n$, where $n_c$ denotes the code word, $A$ the encoding constant, and $n$ the functional value.
Hence, all multiples of $A$ are valid code words. To validate the code word, 
the AN-code congruence in the form $0 \equiv n_c \textrm{ mod } A$ is applied.\looseness=-1

AN-codes limit the functional value to be less than $A$ to preserve the error detection capabilities. The encoding constant is chosen by the designer and defines the redundancy properties of the code.
The minimum Hamming distance between the code words gives a quantitative measure how strong the chosen $A$ is. Finding a good $A$ so far is limited by exhaustive search~\cite{DBLP:conf/wisa/MedwedS09}, but good encoding constants already have been found. Hoffmann~\etal.~\cite{DBLP:conf/hase/HoffmannUDSLS14} call these constants as so-called \textit{Super A}s because their minimum Hamming distance is maximal for a given word width.

Since AN-codes are closed under addition and subtraction (Equation~\ref{eq:an_add_sub}), these operations do not require any modification. Operations like multiplication are supported but require a special correction value.
\vspace{-0.4em}
\begin{equation}
  \label{eq:an_add_sub}
	z_c = x_c + y_c = A \cdot x + A \cdot y = A \cdot (x + y) = A \cdot z
\end{equation}
Fetzer~\etal.~\cite{DBLP:conf/safecomp/FetzerSS09} use this encoding scheme to build an AN-code LLVM compiler, which transforms all operations to the domain of AN-codes, to protect the data processing. However, as discussed by Hoffmann~\etal.~\cite{DBLP:conf/hase/HoffmannUDSLS14}, AN-codes alone is not sufficient because conditional branches are still a single point of failure.

\subsection{Duplication}
\label{sec:duplication}

One way to avoid different schemes for protecting code execution and data is modular redundancy~\cite{Barenghi2010}, where each instruction is duplicated.
After a duplicated instruction, a check is inserted. Conditional branches are protected by replicating the branch multiple times resulting in a comparison tree. However, inducing the same fault multiple times bypasses this protection.
Barry~\etal.~\cite{DBLP:conf/hipeac/BarryCR16} automate this, where they duplicate instructions and store the result always in the same result register and avoid check operations. However, this countermeasure is only suitable in the instruction skip fault model.

\section{Protecting Conditional Branches}
\label{sec:Protected_Conditional_Branches}

A conditional branch consists of two operations: a comparison and a branch. The comparison takes two inputs $x$ and $y$, compares them with a predicate $P$ (e.g.,~$<$), and results in a 1-bit signal indicating if the comparison is true or false. Typically, this signal is part of the CPU flags.
The branch takes this signal and decides how to update the program counter~(PC), which can end up with two different values $PC_1$ and $PC_2$, depending on whether the branch was taken or not. 

\begin{figure}[t]
	\centering
	\includegraphics[width=0.25\textwidth]{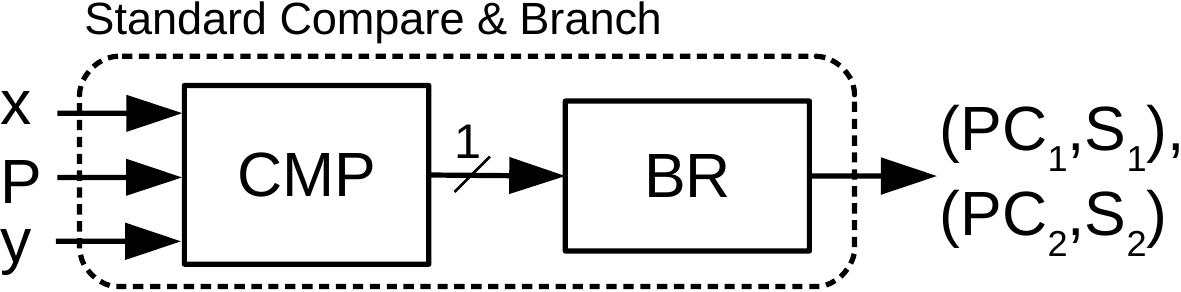}
	\caption{Conditional branch with CFI state.}
	\label{fig:cfi_cond_branch}
\end{figure}

In the presence of a CFI protection mechanism, conditional branches work differently. Again, there is a compare and branch operation as shown in \figurename~\ref{fig:cfi_cond_branch}. However, the CFI protection mechanism contains a dedicated internal state $S$ for each value of the PC, which is updated when executing the conditional branch.
Here, the output of a conditional branch is two different PC values $PC_1$ and $PC_2$ with their corresponding CFI states $S_1$ and $S_2$.\looseness=-1

However, even in the presence of a CFI protection scheme, there are three different error sources, which are not protected and can lead to a wrong execution:

\begin{enumerate}
	\item \textit{Faulting the operands.} Modifications on the branch operands or any data that leads to the comparison can result in a wrongly executed conditional branch.
	\item \textit{Faulting the comparison.} The value deciding whether a conditional branch is taken or not, the condition signal, is a 1-bit signal. An attacker being able to control this signal precisely can change the execution of the conditional branch.
	\item \textit{Faulting the branch.} A fault modifies the execution of the branch such that the branch is taken although the condition value says otherwise or vice versa.
\end{enumerate}
To protect conditional branches, we assume that data and all performed operations on it are encoded redundantly, e.g.,~via AN-codes. 
We generically address the latter two points as follows: first, we use a redundantly encoded condition computation, to ensure the integrity of the condition value. This encoded comparison takes two encoded values $x_{c}$ and $y_{c}$, a comparison predicate $P$, and outputs a redundantly encoded condition result, which Hamming distance is large enough to maintain the same security level throughout the whole conditional branch.
The comparison predicate $P$ does not require redundancy by means of encoding since a different predicate uses a different expected condition value.\looseness=-1
We then use the standard compare and branch mechanism that compares the redundant comparison result with one of the expected condition values.\looseness=-1

\begin{figure}[t]
	\centering
	\includegraphics[width=0.45\textwidth]{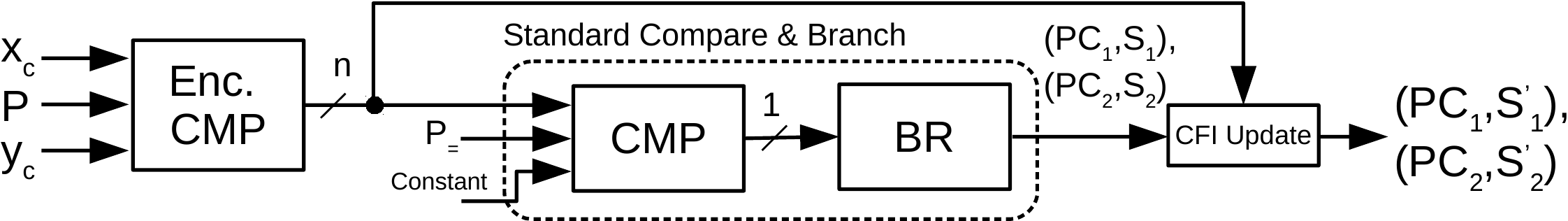}
	\caption{Protected conditional branch with state update and n-bit redundantly encoded comparison.}
	\label{fig:dcfi_cond_branch}
\end{figure}

Without further measure, this introduces an intermediate unprotected 1-bit signal. To mitigate this, we exploit the redundancy of the encoded comparison result and merge this value as part of the CFI state update into the redundancy of the CFI scheme (\figurename~\ref{fig:dcfi_cond_branch}). 
Only if the condition is computed correctly and the branch was executed correctly, the states for $S_1^\prime$ and $S_2^\prime$ are correct. This approach eliminates the single point of failure present in state-of-the-art CFI protection schemes by not relying on a 1-bit condition value but on a redundantly encoded condition value linked with the CFI state.
The comparison is protected by using an encoded comparison operation that yields a redundant result. 
The final conditional branch is protected by linking the redundant condition value with the CFI redundancy. Fault attacks in both cases yield an invalid state $S$, which is detectable.\looseness=-1

Using an encoded comparison operation ahead of an ordinary conditional branch makes this design modular and flexible allowing different encodings with different security levels to be used at various program locations.
The only requirement for the CFI protection is the ability to merge a value into the internal state.
A dedicated conditional branch, which automatically performs the CFI state update and linking the result of the encoded condition value with the CFI redundancy can increase the performance.\looseness=-1

\section{Protected Comparisons with AN-Codes}
\label{sec:AN-Code_Comparisons}

In this section, we discuss a redundant comparison framework which is exploiting the arithmetic properties of AN-codes, which adheres to the interface definition in Equation~\ref{eq:generic_comparison}. The inputs, the internals, and the output are encoded such that there is no single point of failure.
The two possible outputs of the encoded comparison operation should have a Hamming distance larger or equal than a constant $D$, where $D$ denotes the minimum security level in bits of the data encoding and the CFI redundancy.
Furthermore, we want to avoid the all-zero, and all-one condition results because faulting to these values is easier than to others  due to the hardware implementation (e.g.,~the reset line of a register can initialize its value to zero).

\vspace{-1.5em}
\begin{align}
	\label{eq:generic_comparison}
	\text{condition} \leftarrow &\text{EncodedCompare}\left(P, x_c, y_c\right) \\
	& \text{ with condition} \in \{C_1, C_2\} \text{ and} \nonumber\\
	& \text{ Hamming distance}\left(C_1, C_2\right) \geq D \nonumber
\end{align}
AN-codes can be compared using a standard compare instruction. However, this removes all redundancy and results in a 1-bit signal stored inside the CPU. Hoffmann~\etal.~\cite{DBLP:conf/hase/HoffmannUDSLS14} found this issue during fault simulation. Instead, we compute the comparison and preserve the redundancy of the AN-codes avoiding this single point of failure.

To compute the $x_c < y_c$ comparison ($x_c$ and $y_c$ are AN-coded), we start with a subtraction. Based on the sign of this result, we get the information shown in \ref{eq:arith_diff_lt}.
However, we cannot directly use the sign bit because it is not redundant.
The challenging task is performing an entropy compression, where we map the encoded positive difference values to $C_1$, and all encoded negative values to $C_2$.
Additionally, we want to maximize the Hamming distance between $C_1$ and $C_2$ yielding a redundant comparison result.\looseness=-1

\begin{equation}
  \label{eq:arith_diff_lt}
  x_c - y_c
  \begin{cases}
  	\text{positive if} & x_c \geq y_c \\
    \text{negative if} & x_c < y_c
  \end{cases}
\end{equation}
Our approach \textit{arithmetically} computes this entropy compression yielding a comparison result which preserves the redundancy of the AN-code. When looking at the difference in Equation~\ref{eq:arith_diff_lt}, the congruence \mbox{$0 \equiv (x_c - y_c) \text{ mod } A$} is valid because AN-codes are closed under subtraction in a signed representation.
However, when interpreting the AN-code congruence in an unsigned representation, this destroys the congruence for negative differences.
For a positive difference, on the other hand, the unsigned representation does not change anything.
By intentionally destroying the AN-code congruence for negative numbers due to casting to unsigned, we are able to separate the two cases of Equation~\ref{eq:arith_diff_lt} yielding two different values.
Using 32-bit data types, the unsigned interpretation $x_u$ of a signed negative value $x_s < 0$ in the twos-complement representation is computed as $ x_u = 2^{32} + x_s$.
We exploit this property of twos-complement encoded negative numbers for the required entropy compression. First, the difference is cast to an unsigned value. This does not change the difference if it was positive.
Negative values change according to the twos-complement, where the AN-encoded difference becomes invalid. In Equation~\ref{eq:diff_an_code_two_complement}, we show the conversion from the signed AN-code to the unsigned representation for negative values of the difference.
\begin{equation}
  \label{eq:diff_an_code_two_complement}
  (x_c - y_c)_u  = 2^{32} + (x_c - y_c) = 2^{32} + A \cdot (x - y)
\end{equation}
When applying the AN-code congruence to that value by using a modulo operation with $A$, we obtain a dedicated value for the negative difference as shown in Equation~\ref{eq:diff_remainder}.
\begin{equation}
  \label{eq:diff_remainder}
  \left(2^{32} + A \cdot (x - y)\right) \% A = 2^{32} \% A
\end{equation}
The relation described before only holds true for the negative difference. For a positive difference in Equation~\ref{eq:arith_diff_lt}, the AN-code congruence still returns zero. However, as discussed before, having a comparison result that is zero is not favorable. We avoid this zero comparison result for the true case by adding constant $0 < C < A$ to the difference before we compute the remainder (this also changes the comparison result for the false case).

Algorithm~\ref{alg:arithmetic_an_less_than} summarizes how the encoded less-than comparison is computed. The comparison result \textit{cond} holds the value $2^{32} \% A + C$ if $x_c$ is less than $y_c$ or the value $C$ if $x_c$ is larger or equal than $y_c$.
A modification (e.g.,~due to a fault) to the operands such that their AN-code gets invalid results in a different comparison result, making it invalid.\looseness=-1

\begin{algorithm}[t]
	\DontPrintSemicolon
	\KwData{$x_c, y_c \in$ AN-code, $ 0 < C < A$.}
	\KwResult{$cond \in \{C_1,C_2\}$.}
	\Begin{
		diff $\longleftarrow$ (unsigned) $x_c - y_c + C$\;
		cond $\longleftarrow$  diff \% A\;
	}
	\caption{AN-encoded $<$ comparison.\label{alg:arithmetic_an_less_than}}
\end{algorithm}

The same scheme applies to compute a $\leq$, $>$, and $\geq$ comparison by swapping the operands in the first subtraction and swapping the symbols for the true and false case, as summarized for 32-bit data types in \tablename~\ref{tab:an_coded_condition_values}.

\begin{table}[b]
	\centering
	\vspace{-0.7em}
	\caption{Condition values for encoded $<,\leq, >, \geq$ condition values.}
	\label{tab:an_coded_condition_values}
	\begin{tabular}{llll}
		\hline
		Predicate & Subtraction & True Value         & False Value \\ \hline
		$>$       & $y_c - x	_c$ & $2^{32} \% A + C$ & $C$ \\
		$\geq$    & $x_c - y_c$ & $C$                & $2^{32} \% A + C$ \\
		$<$       & $x_c - y_c$ & $2^{32} \% A + C$  & $C$ \\
		$\leq$    & $y_c - x_c$ & $C$                & $2^{32} \% A + C$ \\ \hline
	\end{tabular}
\end{table}

\subsection*{Protected Equal and Not-Equal Condition Computation}
\label{sec:protected_an_eq_operation}

To compute the $=$ and $\neq$ condition, we combine the $\leq$ and $\geq$ condition. 
The $=$ condition is true if both conditions are true and false if only 
$\leq$ is true or $\geq$ is true. 
Both conditions cannot be false at the same time. We combine these conditions using an addition. Using the condition values for $\geq$ and $\leq$ from \tablename~\ref{tab:an_coded_condition_values}, the sum of both true values is $2\cdot C$. 
The false case is the sum of one true and one false case resulting in the condition value $2^{32} \% A  + 2 \cdot C$. 
The algorithm to compute the $=$ or $\neq$ condition is shown in Algorithm~\ref{alg:arithmetic_an_equal}.\looseness=-1
\begin{algorithm}[t]
	\DontPrintSemicolon
	\KwData{$x_c, y_c \in$ AN-code, $ 0 < C < A$.}
	\KwResult{$cond \in \{C_1,C_2\}$.}
	\Begin{
		diff1 $\longleftarrow$ (unsigned) $x_c - y_c$\;
  		diff1 $\longleftarrow$ diff1 + C\;
  		rem1 $\longleftarrow$ diff1 \% A\;
  		diff2 $\longleftarrow$ (unsigned) $y_c - x_c$\;
  		diff2 $\longleftarrow$ diff2 + C\;
  		rem2 $\longleftarrow$ diff2 \% A\;
  		cond $\longleftarrow$ rem1 + rem2
	}
	\caption{AN-encoded $=$ and $\neq$ comparison.\label{alg:arithmetic_an_equal}}
\end{algorithm}

\paragraph{Parameter Selection.}

For the comparison algorithms, we used 32-bit registers and chose $A$ to be $63877$ (a \textit{super}-A according to Hoffmann~\etal.~\cite{DBLP:conf/hase/HoffmannUDSLS14}). This $A$ maximizes the functional value for 16-bit data and has a minimum Hamming distance of \textit{six} between all code words, allowing the code to detect up to 5-bit errors. 
We then chose $C$ such that it maximizes the Hamming distance between the true and false symbol for one comparison.  For the $=$ and $\neq$ comparison we select $C=14991$ and for the $<,\leq,>,\geq$ comparison we select $C=29982$. With both constants, we reach a maximum Hamming distance $D$ of 15-bit between the comparison values.\looseness=-1

\section{Implementation and Evaluation}
\label{sec:Implementation_Evaluation}

\begin{figure}[t]
  \centering
  \includegraphics[width=0.5\textwidth]{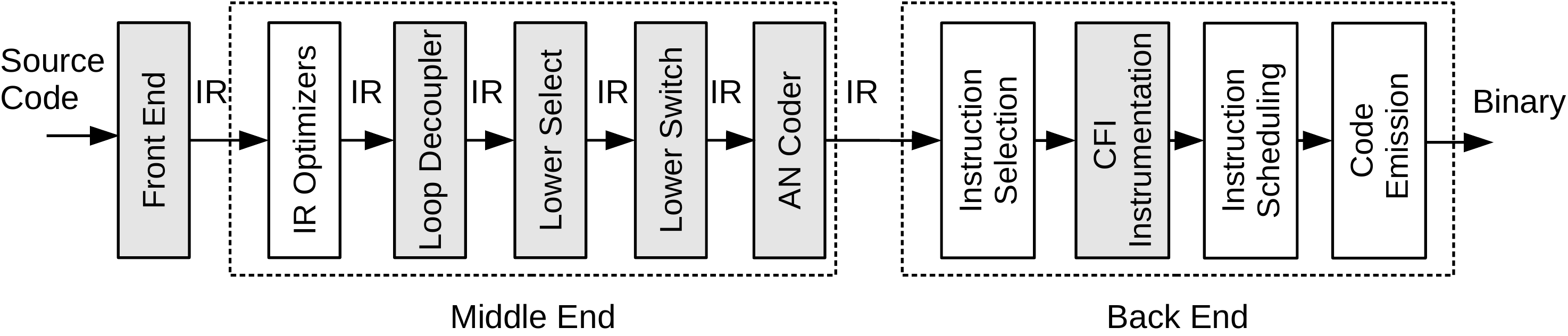}
  \caption{Modified LLVM compiler pipeline. Grey boxes indicate modifications or additions of/to the regular compilation flow.}
  \label{fig:llvm_pass_pipeline}
\end{figure}

We included all transformations to the LLVM compiler (\figurename~\ref{fig:llvm_pass_pipeline}) and evaluated this scheme using an \mbox{ARMv7-M} instruction set architecture~(ISA) simulator.
We use a software-centered GPSA CFI scheme similar to the one in~\cite{DBLP:conf/cardis/WernerWM15}.
The branch protection is purely implemented in software and does not require hardware modifications.\looseness=-1

The compiler front end contains a new function attribute (i.e.,~\verb|protect_branches|) to annotate functions that require protection.
The AN-code instrumentation is performed in the middle end. There, the optimized intermediate representation~(IR) is preprocessed by a custom \textit{Loop Decoupler} pass which separates loop induction variables from the use in arithmetic expressions or memory accesses and a \textit{Lower Select/Switch} pass simplifying the IR for the subsequent \textit{AN Coder}.
The \textit{AN Coder} pass transforms all instructions, which end up in the comparison operation of a conditional branch to the AN-domain.
Moreover, the AN-code based \textit{encoded compare} is added here. Up to this point in the compiler pipeline, all transformations are independent of the target architecture and CFI scheme.
The \textit{CFI Instrumentation} pass in the back end is the only architecture and CFI specific part of this design. It performs the CFI instrumentation and adds the \textit{state updates} to the conditional branches.\looseness=-1

\paragraph{Cost Analysis.}

\begin{table}[b!]
  \centering
  \caption{Qualitative overhead analysis of the building blocks.}
  \label{tab:cost_analysis}
  \begin{minipage}{\textwidth}
    \renewcommand{\thefootnote}{\alph{footnote}}
    \begin{tabular}{c|c|ccc}
    	  \hline
      \multirow{2}{*}{Predicate} & Required   & \multicolumn{3}{c}{Our Prototype}\\
                                 & Operations & Instructions &  Size / B & Runtime / c\footnotemark[1] \\
      \hline \hline
      \cell{$>$\\$\geq$\\$<$\\$\leq$} & \cell{1 +\\1 -\\1 \%} & \cell{1 ADD\\1 SUB\\1 UDIV\\1 MLS} & 12        & 6-16 \\ \hline
       \cell{$=$\\$\neq$}             & \cell{3 +\\2 -\\2 \%} & \cell{3 ADD\\2 SUB\\2 UDIV\\2 MLS} & 26        & 13-33 \\ \hline
    \end{tabular}
    \footnotetext[1]{Division on \mbox{ARMv7-M} requires between 2 and 12 cycles.}
  \end{minipage}
\end{table}

The overhead of our implementation comprises three parts: the cost of computation on encoded data yielding into a branch, the costs of the branch protection scheme, and the costs of the CFI scheme. Given that we solely propose a branch protection, we do not focus on analyzing the cost of the used data protection or CFI scheme. These costs are highly application specific and therefore hard to predict. Still, our evaluation indicates that the expected costs for enforcing CFI and for protecting data values are quite reasonable when mostly requiring arithmetic operations.\looseness=-1

Analyzing the cost of the encoded compare and the state update operations (\tablename~\ref{tab:cost_analysis}) is possible precisely. 
The generic implementation\footnote{Special encoding constants may have optimized implementations but different code properties.} of the proposed 
encoded compare comprises additions, subtractions, and modulo operations.
Every ISA typically supports addition and subtraction, but modulo is not necessarily supported directly and therefore often is more costly. 
With the used \mbox{ARMv7-M} ISA, modulo has to be implemented using a combination of a slow division (UDIV) and a multiply+subtract (MLS) instruction. As a result, depending on comparison predicate, between 12 and 26~bytes memory overhead, and 6-33~cycles runtime overhead is generated for one encoded compare. Hardware support for a fast modulo instruction would considerably reduce this overhead.\looseness=-1

The cost for state updates dependents on the CFI scheme. In the software-centered design, they are implemented using one address load and a store of the comparison result to the CFI unit. These instructions are added to the beginning of the successor basic blocks of the protected conditional branch and introduce 4~bytes code and 4~cycles of runtime overhead per instantiation. An optimized CFI and branch protection design can fully omit these costs.

\paragraph{Benchmarks.}

We use two micro-benchmarks to measure the overhead in terms of runtime and code size. 
These benchmarks (\textit{integer compare} and \textit{memcmp}) test the branch protection in isolation by exercising a single integer equal comparison and a secure memory comparison with 128 elements. 
We compare this overhead with a duplication approach, where we duplicate the conditional branch six times consecutively to have a comparable single bit fault tolerance to the AN-code based implementation (i.e.,~6-bit Hamming distance for the encoded values). 
However, this duplication approach does not protect any data or arithmetic operation leading to the branch opposed to the AN-code based scheme. 
As a macro-benchmark, we implement a fault-protected version secure bootloader, similar to the one in~\cite{atmel2017}. Only programs which feature a valid ECDSA signature over the program's hash get executed. In this example, the memory comparison of the signature verification and all subsequent conditional branches are protected. This mitigates the single point of failure of a secure boot mechanism, which was already a target of fault attacks.\looseness=-1

\begin{table}[t]
	\setlength\tabcolsep{2.5pt}
	\centering
	\caption{Size and runtime overhead of different branch protections.}
  	\label{tab:benchmarks}
	\begin{tabular}{cc|c|cc|cc}
	
  	\hline
    \multirow{2}{*}{Benchmark} & \multirow{2}{*}{Metric} & CFI & \multicolumn{2}{c|}{Duplication} & \multicolumn{2}{c}{Prototype}\\
                             &                         & abs & abs & + / \% & abs & + / \% \\
    \hline \hline
    \textit{integer}         & Size / B                &  12    & 128   & 967 & 86   & 617 \\
    \textit{compare}         & Runtime / c             &  20    & 91    & 355 & 63   & 215 \\ \hline
    \multirow{2}{*}{\textit{memcmp}} & Size / B        &  68    & 272   & 300 & 276  & 306 \\
                             & Runtime / c             & 1689   & 10210 & 504 & 8905 & 427 \\ \hline
    \multirow{2}{*}{\textit{bootloader}} & Size / B    & 17252  & ---   & ---  & 17672  & 2.435\\
                             & Runtime / c             & 51888k & ---   & ---  & 51888k & 0.001\\
    \hline
  \end{tabular}
\end{table}

The costs (\tablename~\ref{tab:benchmarks})
also include the overhead of computing on the AN-encoded values. 
Based on the micro-benchmark results, we observe that the performance in terms of code size and runtime is on par with the duplication approach or even better. 
However, we do not only protect the conditional branch but also protect the data and the arithmetic operations on it. 
When applying this protection mechanism to the protected bootloader, the overhead is neglectable since the crypto implementation dominates code size and runtime. The code size overhead of less than 2.5\% and a neglectable runtime overhead makes this countermeasure applicable to real-world applications.

\section{Security Analysis}
\label{sec:analysis}

To state the security of the countermeasure, we analyze its fault resistance.
If there is a fault on a single location but with multiple bits flipped, the error is transparent and detectable relying on the code properties of the selected $A$~\cite{DBLP:conf/safecomp/FetzerSS09}. For our parameter selection, we can detect up to 5-bit errors in a single word during the calculation. In the final condition result, the error detectability is even higher because only two symbols are valid. At this place, we reach a Hamming distance of 15-bit between the two condition values.\looseness=-1

However, if errors are spread over multiple locations/operations, the fault detection capabilities of the AN-code decrease and the code cannot detect as many bits as before.
To investigate this behavior, we performed a simulation with faults at different locations.
Simulations show that for our parameter selection the error detectability is reduced to 3-bits, arbitrarily placed over all the whole computation of the condition value. 
With four bits flipped over the whole computation of a condition value, the error rate where an attacker can flip the final condition value from true to false or vice versa is 0.0002\,\%, which increases having more bits flipped.\looseness=-1

\ifauthor
\section{Acknowledgement}
\label{sec:Acknowledgement}

This project has received funding from the European Research 
Council (ERC) under the European Union’s Horizon 2020 research and 
innovation programme (grant agreement No 681402)
and by the Austrian Research Promotion Agency (FFG) via the competence center Know-Center, which is funded in the context of COMET – Competence Centers for Excellent Technologies by BMVIT, BMWFW, and Styria.

\fi

\section{Conclusion}
\label{sec:Conclusion}

In this work, we close the gap of unprotected conditional branches in CFI countermeasures in the presence of fault attacks. We eliminate the single point of failure by adding an encoded comparison operation that yields a redundant condition value. Using a standard compare and branch mechanism together with the ability to merge the redundant comparison result with the CFI protection mechanism allows us to protect the execution of a conditional branch.
Our approach is highly flexible allowing us to use different encoded comparison operations based on different encoding schemes with different security properties at different places in the program.
We exploit the properties of arithmetic AN-codes and present novel comparison
algorithms to compute the condition values arithmetically but preserve the redundancy.
We integrated this countermeasure in the LLVM compiler to automatically protect conditional branches. Experimental evaluation shows little overhead to security critical programs such as the signature verification of a secure bootloader making it applicable for real-world usage.

\bibliographystyle{IEEEtranS}  

\bibliography{bibliography.bib}

\end{document}